# Influence of substrate bias on the structural and dielectric properties of magnetron-sputtered $Ba_xSr_{1-x}TiO_3$ thin films

*Tommi Riekkinen,[1] Jan Saijets,[1] Pasi Kostamo,[2] Timo Sajavaara,[3] and Sebastiaan van Dijken[4]*

[1] *VTT Technical Research Centre of Finland, P.O. Box 1000, FI-02044 VTT, Finland*

[2] *Micro and Nanosciences Laboratory, Helsinki University of Technology, P.O. Box 3500, FI-02015 TKK, Finland*

[3] *Department of Physics, University of Jyväskylä, P.O. Box 35, FI-40014 University of Jyväskylä, Finland*

[4] *Department of Applied Physics, Helsinki University of Technology, P.O. Box 5100, FI-02015 TKK, Finland*

**Abstract**

The application of a substrate bias during rf magnetron sputtering alters the crystalline structure, grain morphology, lattice strain and composition of $Ba_xSr_{1-x}TiO_3$ thin films. As a result, the dielectric properties of $Pt/Ba_xSr_{1-x}TiO_3/Pt$ parallel-plate capacitors change significantly. With increasing substrate bias we observe a clear shift of the ferroelectric to paraelectric phase transition towards higher temperature, an increase of the dielectric permittivity and tunability at room temperature, and a deterioration of the dielectric loss. To a large extent these changes correlate to a gradual increase of the tensile in-plane film strain with substrate bias and an abrupt change in film composition.

# 1. Introduction

Perovskite metal-oxides such as $Ba_xSr_{1-x}TiO_3$ (BST) exhibit many exceptional material properties including high dielectric tunability and relatively low dielectric loss. The successful integration of these materials into parallel-plate capacitors with low-loss electrodes is of interest to various tunable device applications in microwave technology. As BST is typically grown in an oxidizing atmosphere at temperatures of 600 – 800 ºC, the choice of bottom electrode is limited to metals that are thermally stable, resistant against oxidation, and act as good seed layers for high-quality BST film growth. State-of-the-art bottom electrodes for high-frequency applications currently consist of relatively thick Pt[1] layers or combinations of Pt and Au[2,3]. As an alternative, a layer transfer method for the fabrication of parallel-plate capacitors was recently demonstrated allowing the use of highly conductive electrodes of Ag and Cu.[4]

Besides the choice of bottom electrode material, the deposition method and growth parameters also greatly influence the microstructure and dielectric properties of BST films. High quality polycrystalline BST capacitors have been grown by magnetron sputtering[5,6], chemical vapor deposition[7,8], sol gel[4,9], pulsed laser deposition[10,11] and other methods and the dependence of the dielectric properties on deposition temperature, growth rate, process pressure, composition, and film thickness has been reported[5,6,8,12,13]. The properties of BST thin films are usually inferior to their bulk analogs due to grain size effects, charged defects, the formation of interface layers, and the build-up of lattice strain.[14,15,16] In this paper, we report on the influence of another deposition parameter that thus far has been neglected, namely the application of substrate bias during magnetron sputtering. It is shown that substrate bias can be used to tailor the dielectric response of BST films in a parallel-plate capacitor configuration. The variation of the dielectric permittivity and tunability, the ferroelectric to paraelectric phase transition temperature, and dielectric loss with substrate bias voltage are mainly explained by variations in film strain and composition.

## 2. Experimental

The BST films were deposited onto thermally oxidized silicon substrates with a 50 nm Ti/100 nm Pt seed layer using rf magnetron sputtering from a $Ba_{0.25}Sr_{0.75}TiO_3$ target in a 3:1 $Ar/O_2$ atmosphere. During BST film growth, the substrate temperature was held at 650 ºC. The magnetron sputtering power was 500 W (1.56 W/cm$^2$) and together with a total sputtering pressure of 2.8 mbar and a target-to-substrate distance of 7.5 cm this resulted in a deposition rate of about 3 – 5 nm/min. The total deposition time was fixed at two hours. To examine the influence of substrate bias during BST film growth we deposited films with 0 V (floating), 20 V, 40 V, and 60 V applied bias voltage. On top of the BST films, 100 nm thick electrodes of Al or Pt were patterned by lift-off lithography to form parallel-plate capacitors.

The crystallographic texture and lattice distortion of the BST films was analyzed using x-ray diffraction (XRD). The surface morphology was inspected by various techniques including optical microscopy, scanning electron microscopy (SEM), and atomic force microscopy (AFM). The BST film thickness was measured using surface profilometry and cross-sectional SEM. Rutherford backscattering spectrometry (RBS) using 800 keV incident He$^+$ ions was utilized to examine the elemental composition of the BST films. The dielectric properties of Pt/BST/Pt parallel plate capacitors were characterized from 20 to 300 K at a frequency of 1 MHz with an Agilent 4294A precision impedance analyzer. The capacitor test structures were connected to the analyzer by Cascade microprobes, whereby the signal probe contacted a well-defined 82x82 µm$^2$ top electrode and the ground probe contacted a much larger area that efficiently coupled to the continuous bottom electrode.

## 3. Results

Figure 1 shows the deposition rate and root-mean-square (rms) surface roughness of the BST films as a function of applied substrate bias during sputtering. The deposition rate decreases gradually with substrate bias and this results in a BST film thickness of 630 nm (0 V), 560 (20 V), 400 (40 V), and 330 nm (60 V). The rms surface roughness, on the other hand, does not depend strongly on substrate bias. The average rms surface roughness amounts to about 12 nm, but this does not mean that the surface and grain morphology evolve independently from the applied substrate bias. In fact, the application of substrate bias results in denser BST films and a rounding of the surface grain morphology as illustrated by the SEM images of Fig. 2. If no bias is applied during rf magnetron sputtering, a loose columnar grain structure evolves with various crystalline orientations and clear facets at the surface. The application of a substrate bias provides additional energy to the growth process and it enhances resputtering from the growing BST film. These effects lead to a dense and nearly void free film, rounding of the evolving grain structure, and a reduction of the deposition rate.

The polycrystalline nature of the BST films is confirmed by the XRD measurements of Fig. 3. BST (100), (110), (111), (211), (310), and higher order reflections can be distinguished for most films. However, the relative contribution of the different crystalline phases changes with applied substrate bias. In particular, the number of (100) and (111) grains increases at the expense of (110) and (211) BST grains. The preferential growth of (100) grains can be explained by the low surface free energy of the densely packed BST (100) plane. The application of a substrate bias provides additional energy to the growth process and this promotes the formation of energetically favourable crystalline orientations. A similar shift from (110) to (100) oriented BST films on a Pt(111) seed layer has been observed with increasing deposition temperature and this result can also be explained by the surface free energy argument[10]. The preferential growth of (111) oriented grains during magnetron sputtering with substrate bias is less obvious due to an overlap between the BST (111) and Pt (111) reflections in the XRD $\theta-2\theta$ scans. However, the intensity of the BST/Pt (111) reflection increases more than twofold when the substrate bias is raised from 0 V to 20 V. This is a clear confirmation that the Pt (111) bottom electrode induces a high degree of BST (111) film texture[11] and that the formation of (111) grains is more efficient when a substrate bias is applied during rf magnetron sputtering. Besides the change in the relative contribution of the different crystalline orientations, we also observe an improvement of the film texture with the application of substrate bias. This is for example illustrated by the evolution of the full width at half maximum (FWHM) of the BST (200) reflection with bias voltage; 0.667° (0 V), 0.442° (20 V), 0,410° (40 V), and 0,411° (60 V).

The growth of BST films on Ti/Pt seed layers at high temperatures normally induces lattice strain. First of all, a difference in the coefficient of thermal expansion (CTE) of the silicon substrate, the thermal $SiO_2$ layer, the Ti and Pt seed layers, and the BST film strains the BST lattice when the substrate is cooled from the deposition conditions (650 °C in our experiments) to room temperature. The CTE values for Si, $SiO_2$, Ti, Pt, and BST are 2.6, 0.55, 8.6, 8.8, and 10.5 x $10^{-6}$ $K^{-1}$, respectively. Cooling after deposition is therefore expected to result in tensile in-plane and compressive out-of-plane BST film strain. Other factors that determine the strain state of the BST film include the lattice misfit with the Pt bottom electrode ($a_{Pt}$ = 0.3924 nm, $a_{BST}$ = 0.3930[11]) and deposition parameters such as substrate temperature, sputtering power and pressure, and substrate bias. Figure 4 summarizes the out-of-plane BST lattice distortion of grains with a (100) and (111) crystalline orientation relative to their bulk values as a function of applied substrate bias. The data of Fig. 4 were extracted from the shift of the BST (200) and (222) reflections to larger angles in the XRD $\theta-2\theta$ scans. The BST film grown without applied bias exhibits a reduced out-of-plane lattice parameter (about -0.3% compared to bulk BST) and thus a tensile in-plane film strain. This agrees qualitatively with the difference in CTE values of the substrate material, seed layers, and BST film. The application of a bias voltage during magnetron sputtering strains the BST lattice further. The out-of-plane lattice parameter is about 0.6% smaller than the bulk BST value at high bias voltages (40 and 60 V) and this additional lattice distortion also increases the tensile in-plane film strain.

The dielectric properties of BST films depend critically on the elemental composition. To determine how the substrate bias influences the film composition we conducted detailed RBS experiments. Figure 5 summarizes the Ba/Sr and (Ba + Sr)/Ti atomic ratios as inferred from data fits to RBS measurements with an uncertainty of ± 5%. At low bias voltages (≤ 40 V), the Ba/Sr ratio is about 1/3 which agrees well with the nominal composition of the BST sputter target. The (Ba + Sr)/Ti ratio, on the other hand, is only about 0.92 in the same films. In some reports, a similar Ti non-stoichiometry has been observed when the BST films were grown at low sputtering pressure.[6,12] As our films were grown in a total Ar/O$_2$ gas pressure of only 2.8 mbar this most likely also explains the relatively larger Ti concentration in our films. An abrupt change in the film composition occurs when the substrate bias voltage is increased from 40 V to 60 V. The Ba/Sr ratio in the BST film increases to about 0.41, which is considerably larger than the ratio of the sputter target. Moreover, the Ti non-stoichiometry almost completely disappears at 60 V substrate bias, i.e. the (Ba + Sr)/Ti ratio equals 1 within experimental uncertainty. Preferential resputtering from the growing BST film most likely explains the observed shift in film composition. During BST growth at high substrate bias the growing film is bombarded with highly energetic ions and this results in an efficient resputtering of the deposition material. The resputtering yield ($Y$) depends on the atomic mass and binding energy. The experimental result of Fig. 5 suggest that $Y_{Ti} > Y_{Sr} > Y_{Ba}$, which qualitatively agrees with the notion that light elements are resputtered more easily than heavy elements.

The BST films undergo a diffuse ferroelectric to paraelectric phase transition at low temperatures as illustrated by the broad maxima of the relative dielectric permittivity versus temperature curves in Fig. 6. The maximum of the dielectric permittivity shifts to higher temperatures if the BST films are sputtered onto a biased substrate. The transition temperatures are 120 K (0 V), 140 K (20 V), 180 K (40 V), and 210 K (60 V). For BST single crystals, the ferroelectric to paraelectric phase transition is considerably sharper and the dielectric permittivity in the paraelectric state is described by the Curie-Weiss law $\varepsilon_r(T)=C/(T-T_C)$, where $C$ is the Curie-Weiss constant and $T_C$ is the Curie temperature. Experimental data on $Ba_xSr_{1-x}TiO_3$ single crystals with different Ba content $x$ have been collected and polynomial fits to this data resulted in $T_C = 42+439.37x-95.95x^2$ and $C = 86000+110000x^2$.[14] Filling in the RBS measured Ba content of our BST films (Fig. 5) into these equations gives expected Curie temperatures of 141 K (0 V), 138 K (20 V), 146 K (40 V), and 161 K (60 V). The resemblance between the calculated Curie temperature based on single crystalline defect-free bulk ceramics and the transition temperature of our defected, polycrystalline, and non-stoichiometric BST thin films is remarkable and most likely somewhat coincidental. The observed shift of the ferroelectric to paraelectric phase transition to higher temperature with increasing substrate bias, however, is at least partly explained by an increase of the Ba content due to preferential resputtering of Ti and Sr.

Figure 7 shows the electric field dependence of the relative dielectric permittivity of the Pt/BST/Pt parallel-plate capacitors at 1 MHz. The zero-field dielectric permittivity increases non-monotonically with applied substrate bias during rf sputtering of the BST film. For a substrate bias of up to 40 V the increase of the dielectric permittivity is rather modest (from 350 to about 410) and these values are lower than 600, the expected relative permittivity for bulk BST crystals with the same composition. Increasing the substrate bias to 60 V drastically enhances the zero-field permittivity to 780, which in this case is within experimental error of the calculated bulk value of about 700. The dielectric tunability defined as $(\varepsilon(0)-\varepsilon(E))/\varepsilon(0)$ also increases with substrate bias voltage. For the non-biased BST film, the tunability at an electric field of 3 MV/cm is 35% and it increases to 55% for a substrate bias of 60 V (see inset of Fig. 7).

The inverse of the dielectric loss or so-called quality factor of the different BST capacitors at 1 MHz is summarized in Fig. 8. The BST films grown without substrate bias and a bias of 20 V exhibit a large quality factor of about 200. Interestingly, the electric-field dependence of these two capacitors is different. A leakage current decreases the quality factor of the BST film grown without substrate bias above 1 MV/cm. No such effect is measured when a 20 V bias is used during BST sputtering. Instead, the quality factor increases with applied electric field and this is a clear indication of the quality and relatively low defect density of this BST film. This observation is also confirmed by measurements of the breakdown field, which is largest for BST films grown with a substrate bias of 20 V. The quality factor decreases for higher substrate bias. This is most likely due to defect structures that are created by high energy ion bombardment of the BST film during rf magnetron sputtering at elevated substrate bias.

**Discussion**

The application of substrate bias during rf magnetron sputtering of BST films considerably alters the crystalline texture, surface morphology, strain state, composition, and dielectric properties. With increasing substrate bias voltage the BST film texture gradually changes in favour of (100) and (111) crystalline orientations, the film density increases, the grains become more rounded, the tensile in-plane film strain enhances, and the atomic ratios change in favour of Ba and at the expense of Ti. The most drastic changes in the dielectric response of the Pt/BST/Pt parallel-plate occur when the substrate bias is increased from 40 V to 60 V. This modification correlates with an abrupt increase of the Ba/Sr and (Ba + Sr)/Ti ratios as inferred from the RBS measurements. The higher Ba content of the films grown with a 60 V bias does at least partly explain the shift of the ferroelectric to paraelectric phase transition to higher temperatures and the related increase of the relative dielectric permittivity and dielectric tunability at room temperature. Besides, the change from Ti excess in BST films grown with 0 V, 20 V, and 40 V substrate bias to a stoichiometric ((Ba + Sr)/Ti ≈ 1) BST film at 60 V is also expected to influence the dielectric properties. It has been previously reported that films with a low (Ba + Sr)/Ti ratio exhibit low dielectric permittivity, high quality factors, and low dielectric tunability.[6,8,12] This agrees well with our experiments where an increase of the (Ba + Sr)/Ti ratio at substrate bias of 60 V correlates with an increase of the relative dielectric permittivity to a bulk-like value of 780 at room temperature, an increase of the dielectric tunability to 55%, and a decrease of the quality factor. Also, the large discrepancy between the relative dielectric permittivity of the BST films grown at lower substrate bias (350 – 410) and the calculated bulk value for the measured Ba content of $x = 0.24$ (about 600) can be explained by an excess of Ti in the BST films. As stated earlier, the most likely cause of a reduction of Ti and an enhancement of Ba content at high substrate bias is preferential resputtering from the BST film due to highly energetic ion bombardment during growth. The increase of the (Ba + Sr)/Ti ratio at a substrate bias of 60 V could also partly explain the lower quality factor of this BST film. In addition, the bombardment with high energy ions during deposition is expected to create a variety of other structural defects that contribute to the dielectric loss.

While the correlation between film composition and the dielectric properties of Pt/BST/Pt parallel-plate capacitors is apparent in our experiments, the influence of film strain on the dielectric permittivity, tunability, and loss is somewhat less pronounced. Other groups have reported an increase of the ferroelectric to paraelectric phase transition temperature and a decrease of the maximum dielectric permittivity with increasing mechanical strain for BST films.[7,17,18,19] This qualitatively agrees with the data for 0 V, 20 V, and 40 V substrate bias in Fig. 6. Hence, the additional distortion of the out-of-plane lattice from -0.3% (0 V) to -0.6% (40 V) and thus an increase of the tensile in-plane film strain does at least partly explain the +60 K shift of the transition temperature in these samples. Moreover, as strain moves the maximum of the relative dielectric permittivity towards higher temperatures in the BST films grown with 20 V and 40 V bias, it also increases the permittivity and the dielectric tunability at room temperature. This is clearly illustrated by the data of Figs. 6 and 7. The influence of lattice strain on the dielectric properties of the Pt/BST/Pt parallel-plate capacitor grown with a substrate bias of 60 V is blurred by effects due to the abrupt change in composition as discussed in the previous paragraph.

**Conclusions**

Dielectric properties such as permittivity, tunability, and quality factor of magnetron-sputtered BST films are significantly affected by the application of a substrate bias during deposition. The main effects are due to a bias-induced change in the film composition at very high voltage (60 V) and a gradual increase of the tensile in-plane film strain with substrate bias. These compositional and structural changes do shift the ferroelectric to paraelectric phase transition to higher temperature, increase the dielectric permittivity and tunability at room temperature, and ultimately deteriorate the quality factor of Pt/BST/Pt parallel-plate capacitors. Based on our results, the performance of paraelectric and ferroelectric thin films can be improved by the application of a small to moderate substrate bias voltage ($\approx$ 20 V) during growth. At more elevated bias voltages, the higher energy growth dynamics produces a more defected film and an enhanced dielectric loss. Tailoring of dielectric properties using substrate bias during magnetron sputtering opens a new route to the fabrication of tunable microwave devices.

The authors acknowledge Hannu Hakojärvi for assistance with the low temperature dielectric measurements and Tomi Mattila for valuable discussions.

**Figure Captions**

Fig. 1. Deposition rate and rms surface roughness as a function of bias voltage.

Fig. 2. SEM images of the surface morphology for a BST film grown (a) without substrate bias and (b) with a substrate bias of 40 V.

Fig. 3. $\theta$-$2\theta$ XRD scans of Pt/BST/Al parallel-plate capacitor structures grown with different substrate bias during BST magnetron sputtering.

Fig. 4. Out-of-plane lattice distortion relative to a bulk value of $a_{BST}$ = 0.3930 nm for BST films grown with different substrate bias.

Fig.5. Influence of substrate bias on the Ba/Sr and (Ba+Sr)/Ti ratios of magnetron-sputtered films from a $Ba_{0.25}Sr_{0.75}TiO_3$ target. The lines indicate the Ba/Sr and (Ba+Sr)/Ti ratios of the sputter target.

Fig. 6. Normalized capacitance of Pt/BST/Pt parallel-plate capacitors as a function of temperature. The data was collected at 1 MHz and the capacitance was normalized to its value at 300 K.

Fig. 7. Electric-field dependence of the relative dielectric permittivity for BST films grown with different substrate bias. The data was collected at 1 MHz.

Fig. 8. Quality factor of Pt/BST/Pt parallel-plate capacitors at 1 MHz as a function of applied electric field and substrate bias.

Figure 1

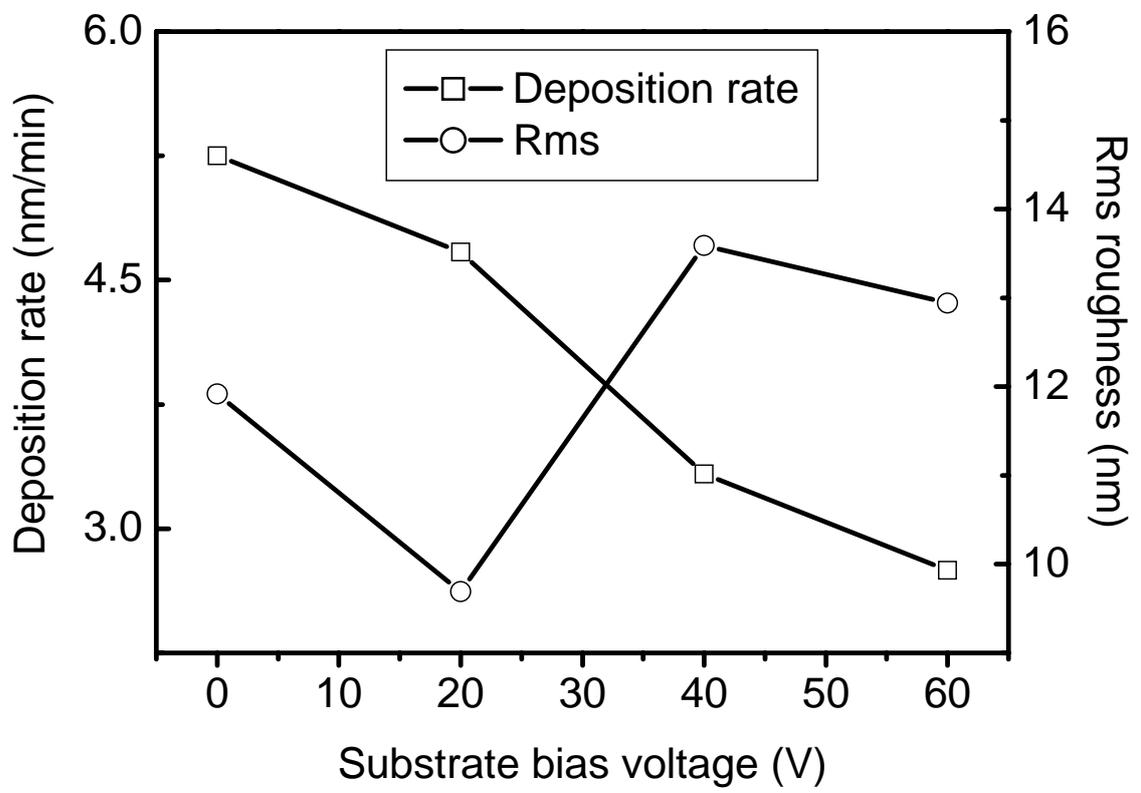

Figure 2

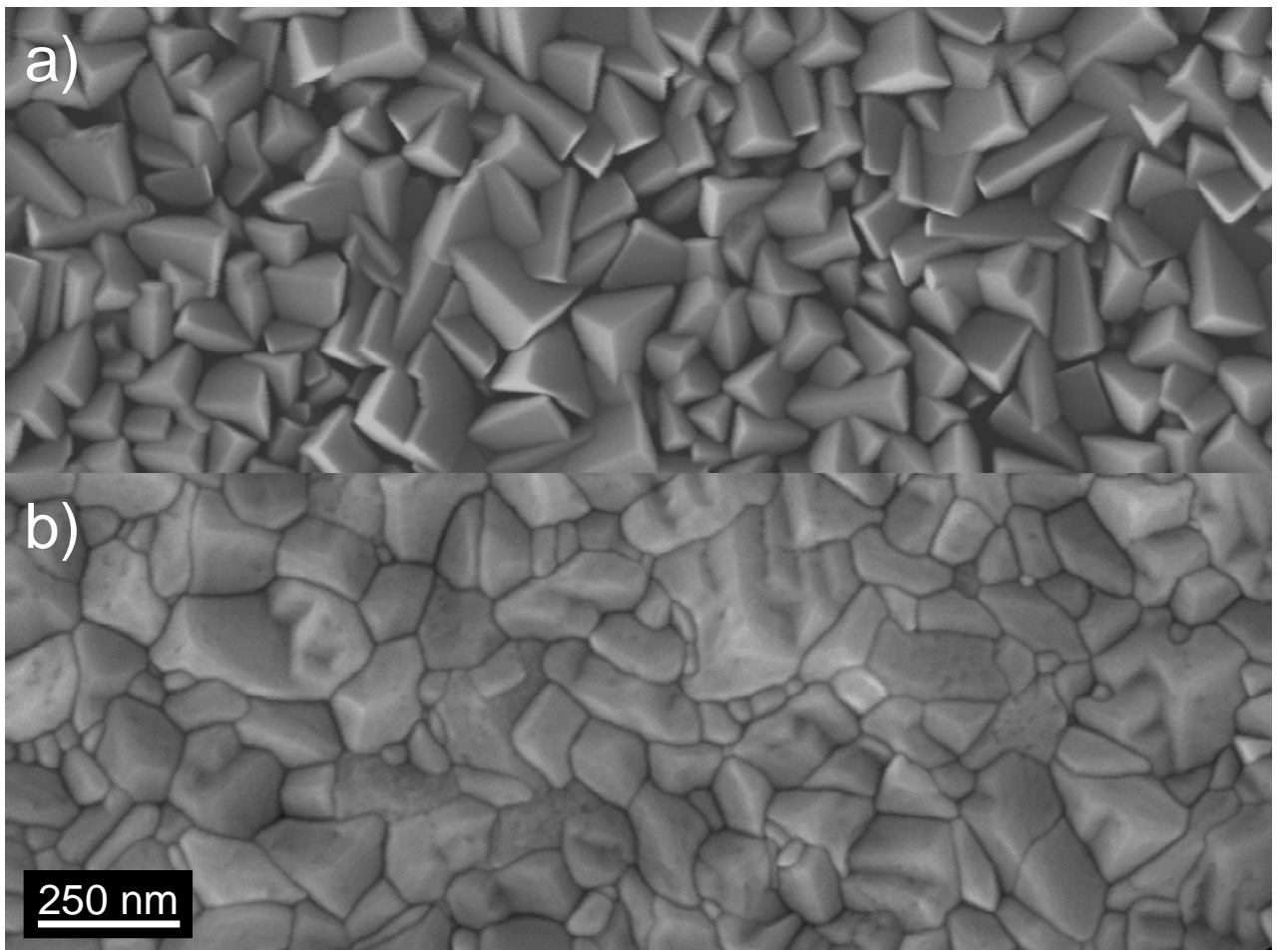

Figure 3

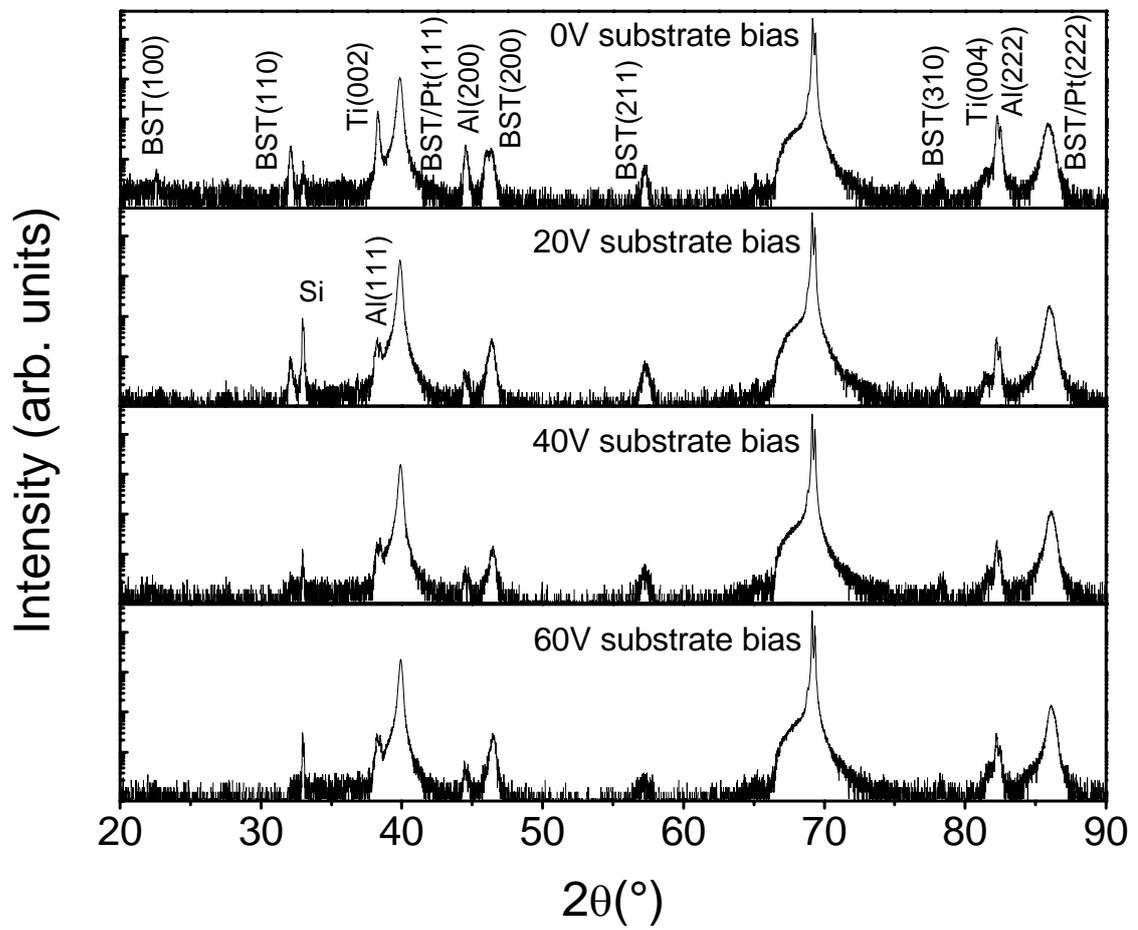

Figure 4

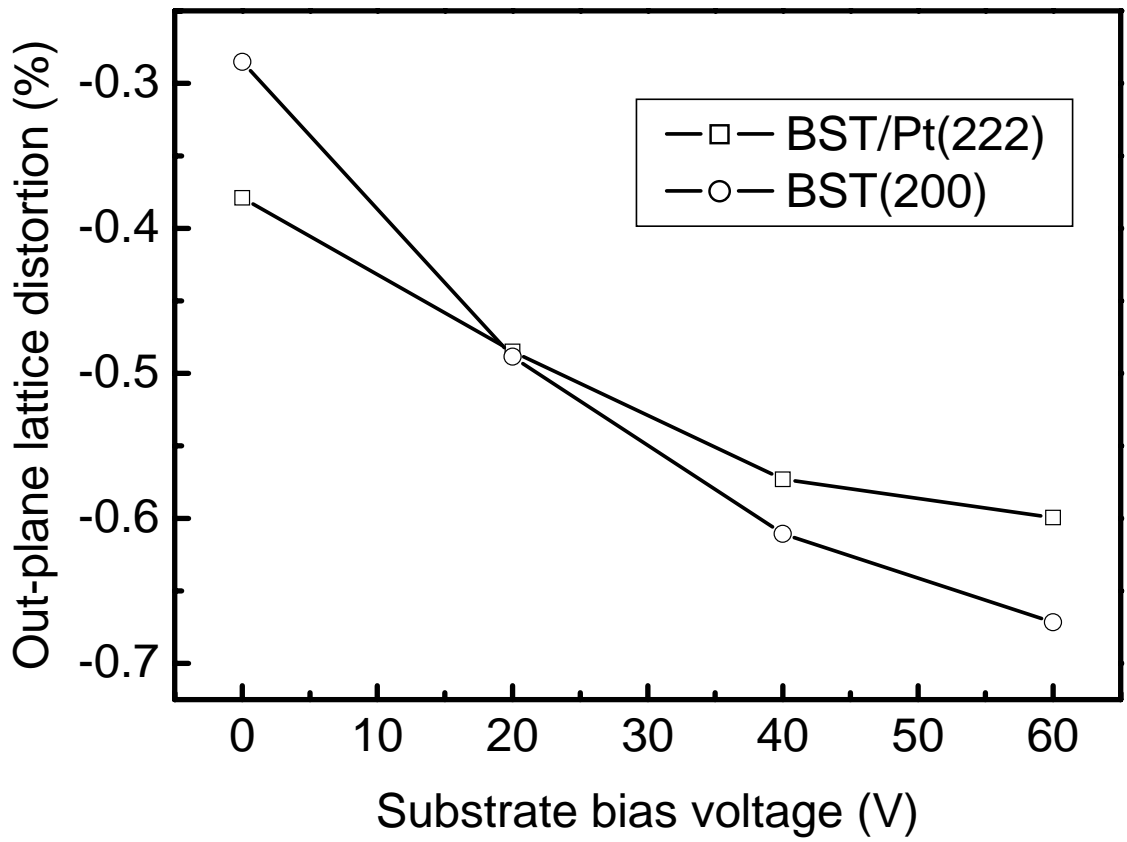

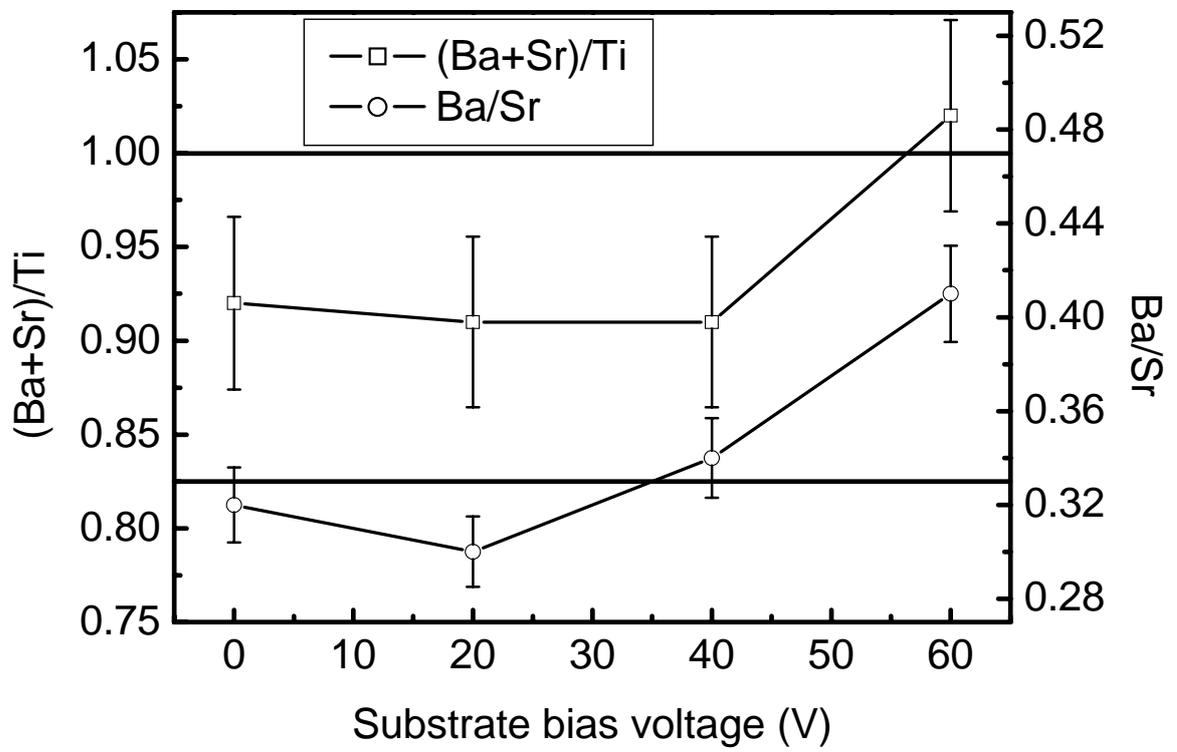

Figure 6

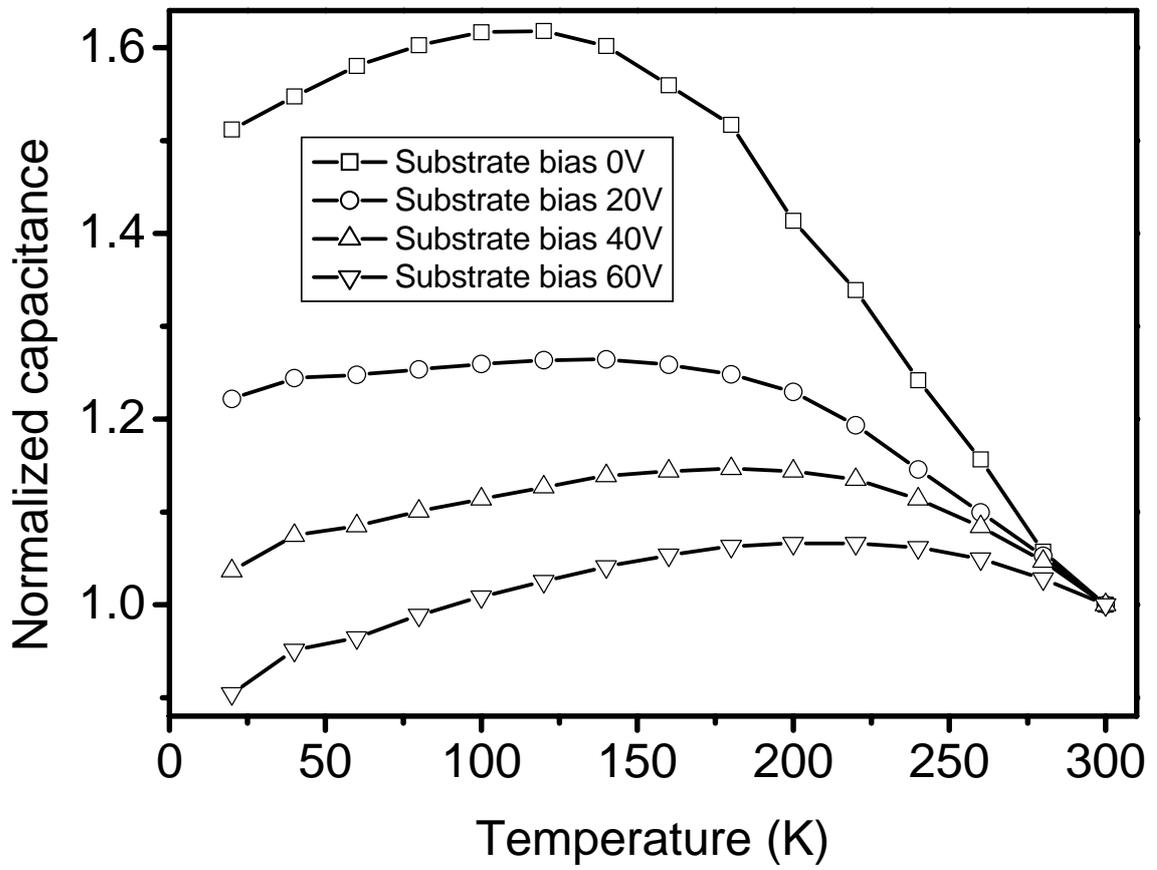

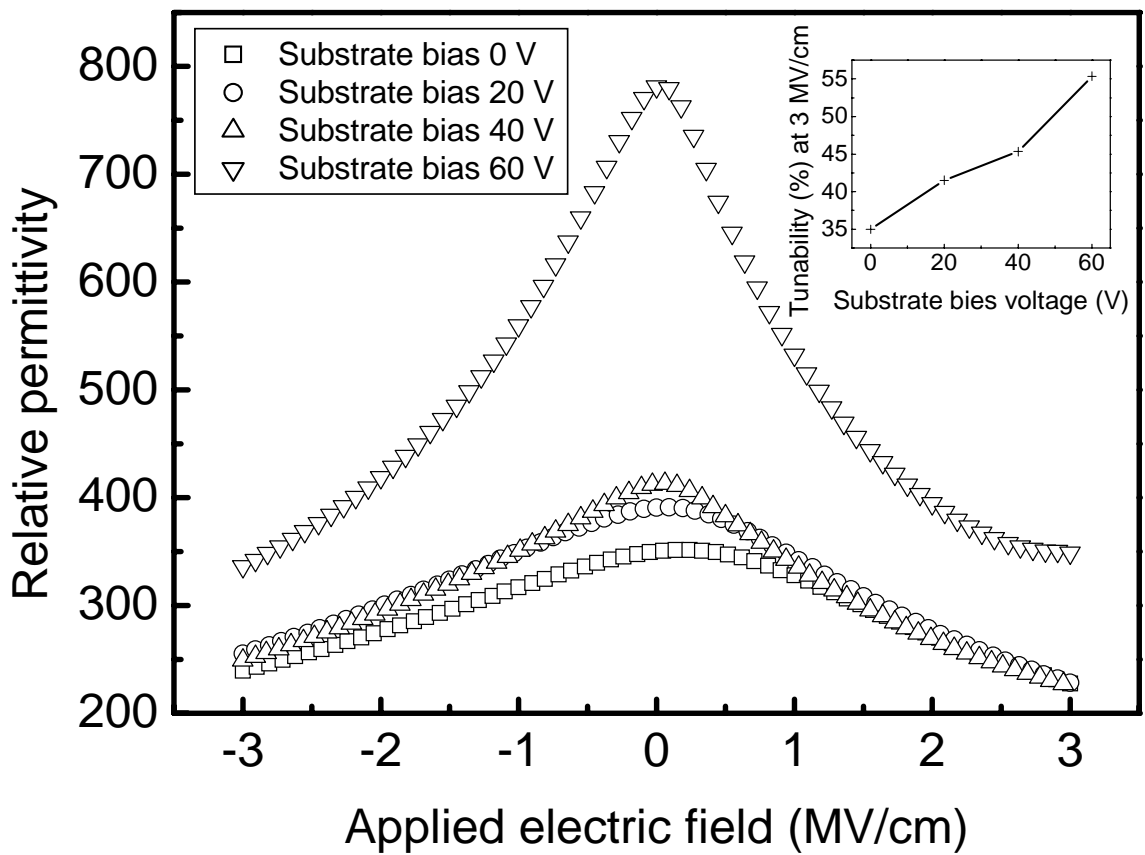

Figure 8

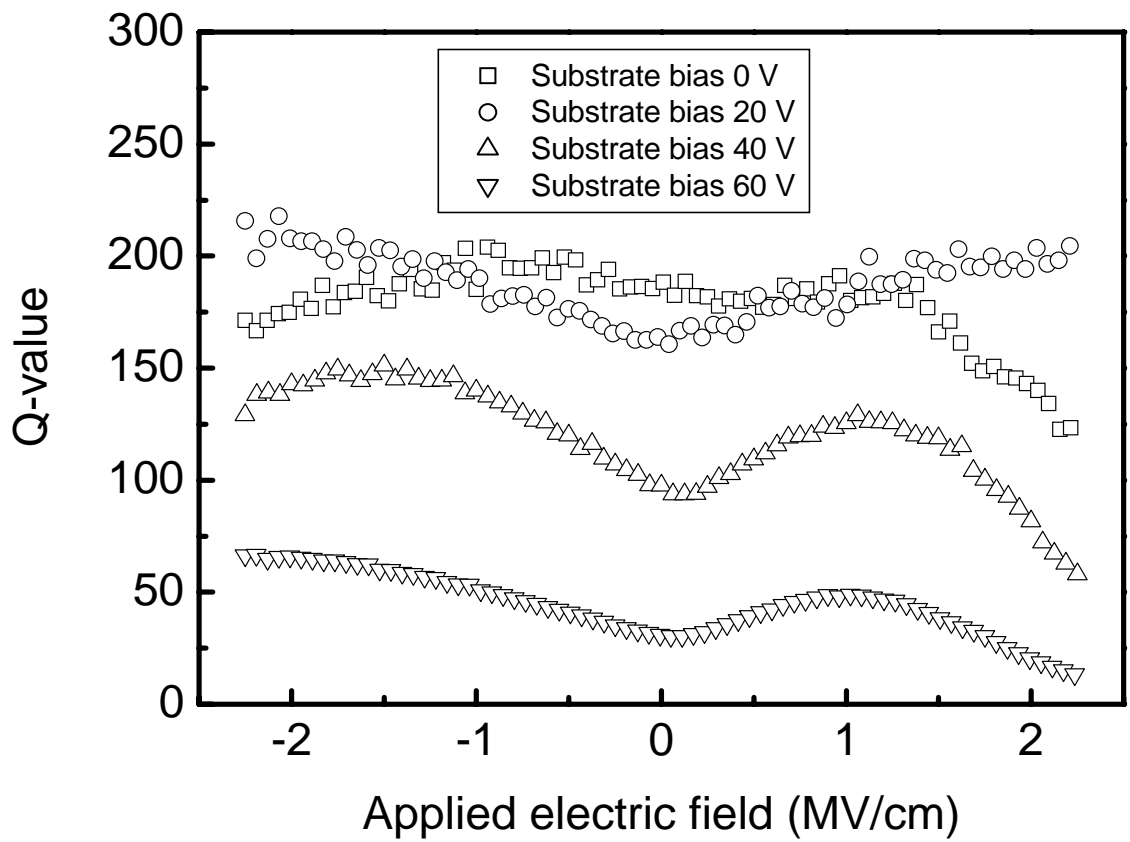